\documentclass{elsart}
\usepackage{epsfig}
\usepackage{amssymb}
\usepackage{color}

\newcommand{\1}{{1\hspace{-3pt} \rm I}}
\begin{document}
\begin{frontmatter}
\title{
$ t $-channel unitarity and photon cross sections}
\author{J.R. Cudell},
\ead{JR.Cudell@ulg.ac.be}
\author{E. Martynov\corauthref{cor1}},
\corauth[cor1]{on leave from Bogolyubov Institute for Theoretical Physics,
Kiev.}
\ead{E.Martynov@guest.ulg.ac.be}
and \author{G. Soyez}
\ead{G.Soyez@ulg.ac.be}
\address{Universit\'e de Li\`ege, B\^at. B5-a, Sart Tilman, B4000 Li\`ege,
Belgium}
\begin{abstract}
We analyse the consequences of $ t $-channel unitarity for the photon hadronic
cross sections. We show that $\gamma^{(*)}p$ and $\gamma^{(*)}\gamma^{(*)}$
elastic amplitudes must include the complex-$j$ plane singularities associated
with hadronic elastic amplitudes, and give the characteristics of possible
new singularities.
We also show that several different models of hadronic total cross sections
can be used to predict the LEP data from the HERA measurements.
\end{abstract}
\end{frontmatter}
\section{Introduction }
\label{intro}
The DIS and total cross section data \cite{H1,ZEUS,total} from HERA
have opened
new avenues in our understanding of strong interactions, and models
\cite{DoLa,DM,CS} now exist which provide a unified description
of $ \gamma p $
interactions for  photon virtualities ranging from
$ Q^{2}=0 $ to $ Q^{2}=30000$~GeV$^{2} $. The theoretical situation
is nevertheless not clear.

Indeed, a wide range of data can be described \cite{DGLAP1,DGLAP2,DGLAP3}
for $ Q^{2}\geq 2 $ GeV$ ^{2} $
by the DGLAP evolution \cite{DGLAP}. Several theoretical
questions need however to be addressed in this context. Firstly, the evolution
is leading twist, and hence one should remove higher-twist contributions
from the data before one uses the DGLAP equation. Secondly, the evolution
introduces extra singularities in the complex-$ j $ plane at $ j=1 $.
These singularities start to appear at the arbitrary factorisation
scale $ Q_{0} $, and their re-summation leads to an essential singularity.
No trace of it is however present in soft cross sections. Finally,
DGLAP \cite{DGLAP} evolution should be replaced at small $ x $ by the BFKL
re-summation \cite{BFKL}. The latter does not lead to an essential singularity
in the complex-$ j $ plane, but unfortunately it does
not seem to be stable against next-to-leading order corrections \cite{NLO}.

Given these problems, Donnachie and Landshoff have proposed to use
the soft pomeron as a higher-twist background to be subtracted from
the evolution, while a new simple pole, the ``hard pomeron''\cite{hardpom},
would reproduce the DIS data. Furthermore, they have shown \cite{DoLa}
that this new singularity evolves according to DGLAP, provided that
one removes the $j$-plane
singularities induced by DGLAP evolution, and keeps only their effect on
the hard
pomeron residue.
Again the question arises whether such a new pole should be present
in total cross sections and whether it is perturbative or not \cite{clms}.

Finally, we have shown that in fact no new singularity is needed to
reproduce the DIS data \cite{DM,CS}, provided that one assumes a
logarithmic behaviour of cross sections as functions of $ \nu $.
Double or triple poles at $ j=1 $ provide such a behaviour, and
enable one to reproduce all soft and hard $ \gamma p $ data within
the Regge region.

How to bridge the gap between those models and QCD remains a challenge,
as the description of the proton, being non-perturbative, remains
at best tentative. However, LEP has now provided us with a variety
of measurements of the $ \gamma \gamma  $ total cross sections, for
on-shell photons, and of $ F_{2}^\gamma $ for off-shell ones \cite{L3,OPAL}.
One may hope that this will be a good testing ground for perturbative
QCD \cite{BL}, and that these measurements will provide guidance for
the QCD understanding of existing models. Hence it is important to
build a unified description of all photon processes, and to explore
where perturbative effects may manifest themselves. The natural framework
for such a program is the ``factorisation theorem'' of the analytic
$ S $ matrix, which relates $ \gamma \gamma  $, $ \gamma p $
and $ pp $ amplitudes. This theorem is based on $ t $-channel
unitarity, \emph{i.e}. unitarity in the crossed channel, and in the
case of simple poles one obtains the factorisation of the
residues at each pole. For more general analytic structures, one obtains
more complicated relations, which we shall spell out in Section 2.

Furthermore, a relation between $ \gamma \gamma  $ and $ \gamma p $
processes may be of practical use as some of the measurements have
big systematic uncertainties. As it is now well known \cite{MCstudy}, the
LEP measurements are sensitive to the theoretical Monte Carlo used
to unfold the data, leading to rather different conclusions concerning
the energy dependence of the data. This problem is manifest in the
case of total cross sections, where the unfolding constitutes the
main uncertainty. In the case of HERA data, the measurement of the
total cross section also seems to be affected by large
uncertainties. Again, a joint study of both processes could help constrain
the possible behaviours of these cross sections.

To decide whether new singularities can appear in $ \gamma p $
and $ \gamma \gamma  $ scattering, one must first recall why singularities
are supposed to be universal in hadronic cross sections. The original
argument \cite{factorisation,multifact}, which we review in Section 2, used
analytic continuation of
amplitudes in the complex-$j $ plane from one side of a 2-particle
threshold to the other, and considered only the case of
simple poles. It leads to the conclusion that these poles must be
universal, and that their residues must factorise.

We show in section 2 that
one can generalise this formula for complex-$ j $ plane
amplitudes, and obtain a formulation
which is valid no matter what the singularity is, and
which leads to consequences similar to factorisation. Indeed, the original
conclusion that singularities had to be simple poles was obtained for
reggeons having resonances on their trajectories. The pomeron case may be
more complicated. For instance, it is possible that no resonances
are present if the real part of the trajectory never reaches integer
values because of non linearity. It is also possible to imagine
situations with harder singularities, such as double or triple poles, or
models involving simple poles
colliding at $t=0$. We want to stress here that our goal is not to decide
theoretically between these possibilities by solving the unitarity
equations, but only to provide a relation between various amplitudes.

We show in
the third section that such a formula may be extended to photon
cross sections for any value of $ Q^{2} $, but that new singularities
may be present in photon cross sections.
If we assume as in \cite{DM,CS} that no other singularity
is present in DIS, stringent constraints come from the positivity
requirement for $ \gamma \gamma  $ total cross sections and $F_2$. We show
that it is possible to obtain a good fit to all photon
data for $Q^2<150$ GeV$^2$
by using either double- or triple-pole parametrisations. For total cross
sections, no extra singularity seems to be needed,
whereas we must introduce the singularities associated with the box
diagram at high $Q^2$.
We conclude this study by
outlining its consequences on the evolution of parton distributions and
on the possibility of observing the BFKL pomeron.
\newpage
\section{$\bf t$-channel unitarity in the hadronic case}
\subsection{Elastic unitarity}
We start by considering the amplitudes for three related processes:\\
\begin{center}
\epsfig{file=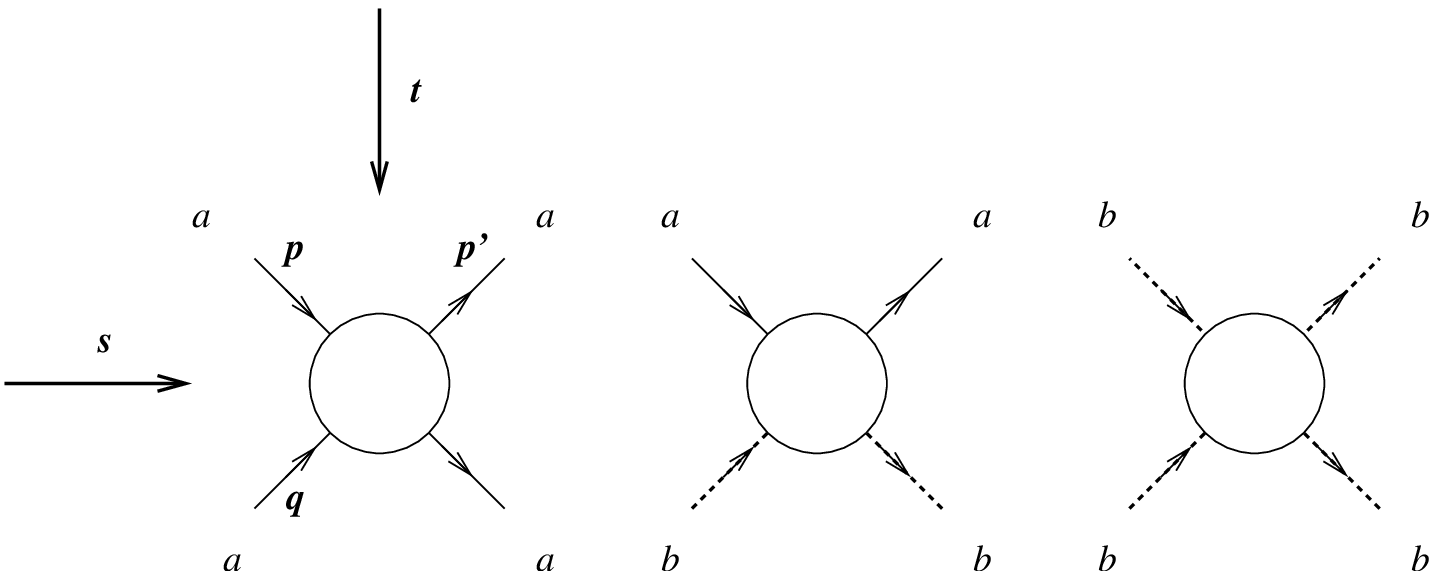,width=11cm}
\end{center}
We shall refer to the momenta of the incoming particles as $p$ and $q$, and to
the outgoing momentum of $a$ as $p'$.
We use the Mandelstam variables $s=(p+q)^2$ and $t=(p-p')^2$.
In the $ s $ channel, these diagrams describe the processes
$ aa\rightarrow aa $,
$ ab\rightarrow ab $, and $ bb\rightarrow bb $. The continuation
of these amplitudes to the $t$ channel
describes the processes $ a\overline{a}\rightarrow a\overline{a}
$,
$ a\overline{a}\rightarrow b\overline{b} $, $ b\overline{b}
\rightarrow b\overline{b} $.

We shall write \( A_{ab}(l,t,m_{a},m_{b}) \) for the $t$-channel partial-wave
elastic amplitude for the process \( a+b\rightarrow a+b \), and denote by
the superscript
$(1)$ the physical-sheet amplitude, and by the superscript $(2)$
its analytic continuation
round a \( c\bar c \) threshold branch point and back to the same value of \( t
\)
(see Fig.~1).
\begin{figure}
\centering{\resizebox*{0.5\columnwidth}{!}{\includegraphics{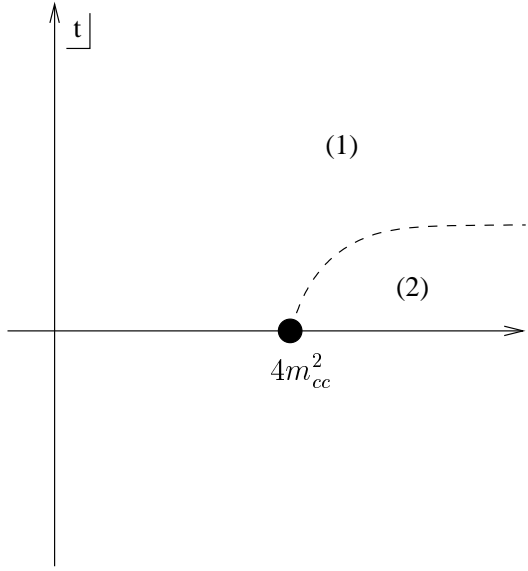}}}
\caption{The amplitude and its continuation in the complex-$t$ plane
around a $c\bar c$ threshold.}
\end{figure}

Unitarity and analyticity of the amplitude then impose the
following relation on the discontinuity through the threshold \cite{DDLN}
\footnote{One
can derive these relations from the unitarity of the $S$ matrix \cite{ELOP}
and we sketch such a derivation in Appendix 1.}:
\begin{equation}
\label{tch}
A^{(1)}_{ab}-A^{(2)}_{ab}=\rho _{c}(t)A^{(1)}_{ac}A_{cb}^{(2)}=\rho
_{c}(t)A^{(2)}_{ac}A_{cb}^{(1)}
\end{equation}
with $\rho _{c}(t)=\sqrt{{t-4m_c^2}\over t}$.

In the case of hadrons, this leads to a closed system of equations,
as the initial
and final states contribute themselves to the thresholds.
For instance, we can consider three coupled
equations for protons and pions, across the \( \pi\pi \) threshold (see
Fig.~2):
\begin{figure}
\centering{\resizebox*{0.5\columnwidth}{!}{\includegraphics{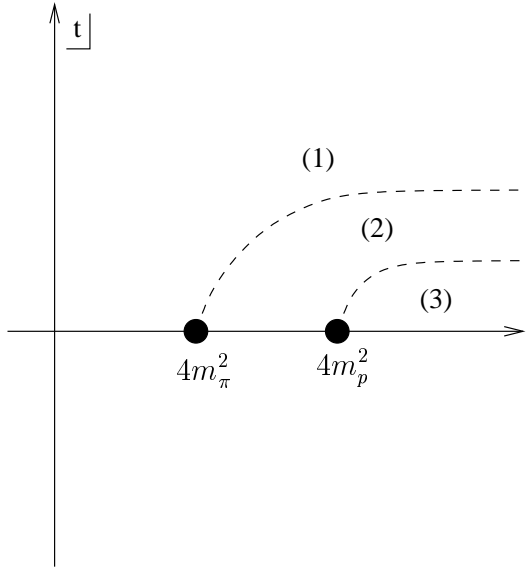}}}
\caption{The amplitude and its continuation
around the $\pi\pi$ and $pp$ branch-points.}
\end{figure}

\begin{eqnarray}
\label{pp}
A^{(1)}_{pp}-A^{(2)}_{pp}&=&\rho _{\pi}(t)A^{(1)}_{p\pi}A_{\pi
p}^{(2)},\nonumber\\
A^{(1)}_{p\pi }-A^{(2)}_{p\pi }&=&\rho _{\pi}(t)A^{(2)}_{p\pi
}A_{\pi\pi}^{(1)},\\
A^{(1)}_{\pi \pi }-A^{(2)}_{\pi \pi }&=&\rho _{\pi}(t)A^{(1)}_{\pi\pi
}A_{\pi\pi }^{(2)}.
\nonumber
\end{eqnarray}

{}From (\ref{pp}), we can write
\begin{equation}
\label{a2pp}
A^{(1)}_{\pi\pi}=\frac{A^{(2)}_{\pi\pi}}{1-\rho _{\pi}(t)A^{(2)}_{\pi\pi}}.
\end{equation}
We see that if \( A^{(1)}_{\pi\pi} \) has a singularity at \( j=\alpha (t) \),
then \( A^{(2)}_{\pi\pi} \) cannot have that singularity, which must be shifted
at
some other value when one goes around the cut. For instance, near
a simple pole at $j=\alpha$,
one has $A^{(1)}\approx {K\over j-\alpha}$, and
$A^{(2)}\approx {K\over j-\alpha+\rho_{\pi} K}$.

Equation (\ref{pp}) can be conveniently written in matrix form:
\begin{equation}
T^{(1)}-T^{(2)}=T^{(1)}{R_\pi}T^{(2)}\label{onet}
\end{equation}
with
\begin{equation}
T^{(i)}=\left( \begin{array}{cc}
A^{(i)}_{\pi \pi } & A^{(i)}_{\pi p}\\
A^{(i)}_{p\pi } & A^{(i)}_{pp}
\end{array}\right)\quad {\rm for}\ i=1,2 \label{T}\end{equation}
and the threshold matrix
\[
R_\pi=\left( \begin{array}{cc}
\rho _{\pi } & 0\\
0 & 0
\end{array}\right). \]

To obtain the most constraining set of equations, one then writes
the discontinuities of $A^{(2)}$ across the $pp$ threshold (the amplitudes
across this threshold are denoted by a superscript (3) and $T^{(3)}$ is defined
according to (\ref{T}) with $i=3$). This
gives, using the same notations as before:
\begin{equation}
T^{(2)}-T^{(3)}=T^{(2)}R_p T^{(3)}\label{twot}
\end{equation}
with
\[
R_p=\left( \begin{array}{cc}
0 & 0\\
0 & \rho_p
\end{array}\right). \]
Putting Eqs. (\ref{onet}, \ref{twot}) together, one then gets:
\begin{eqnarray}
T^{(1)}-T^{(3)}&=&T^{(1)}R_\pi T^{(2)}+T^{(2)}R_p
T^{(3)}\nonumber\\
&=&T^{(1)}(R_\pi+R_p)T^{(3)}
+T^{(1)}R_\pi(T^{(2)}-T^{(3)})\nonumber\\
& &\phantom{T^{(1)}(R_\pi+R_p)T^{(3)}}
+(T^{(2)}-T^{(1)})R_p T^{(3)}\nonumber \\
&=&T^{(1)}RT^{(3)}\label{sol1}
\end{eqnarray}
with $R=R_\pi+R_p$.
Note that $T^{(3)}={T^{(1)}}^\dagger$.

One can then solve the equation for $T^{(1)}$ to obtain
\begin{equation}
T^{(1)}\left[1-RD\right]=D
\label{solut}
\end{equation}
with $D={T^{(1)}}^\dagger$.
This situation generalises that of Eq.~(3): $T^{(1)}$ is
a function of ${T^{(1)}}^\dagger$ and its singularities
cannot come from singularities in the
right-hand side of Eq.~(\ref{solut}), because they are exactly matched
by corresponding factors in the left-hand side. Note that this is why
we needed to consider both thresholds, as otherwise some amplitudes would
not be present in the r.h.s. of (\ref{sol1}).

Hence \emph{
the amplitudes $A^{(1)}_{\pi\pi}$, $A^{(1)}_{pp}$ and $A^{(1)}_{\pi p}$ have
common
singularities}
which
can only come from zeroes $ z_{m} $ of the determinant of the matrix
in brackets in the left-hand side of (\ref{solut}):
\begin{equation}
\Delta =\det \left(1-R D\right) =0\; {\rm for}\; j=z_{m},\
m=1,...,N.\end{equation}
The determinant can then be written
\begin{equation}
\Delta=\prod\limits_{m=1}^N \Delta_m
\end{equation}
with $\Delta_m=0$ at $j=z_m$ and $\Delta_m\neq 0$ at $j=z_k$ if $k\neq m$.
$A^{(1)}_{il}$ (where $i$ and $l$ stand for $p$ or $\pi$) then becomes
\begin{equation}
A^{(1)}_{il}=\sum_{m=1}^N \left({a_{il;\, m}\over \Delta_m}\right)
+F_{il}
\label{sum}
\end{equation}
where $F_{il}$ and $a_{il;\, m}$ are functions of $j$ finite at each $z_m$.

This is the basis of the complex-$ j $-plane factorisation of the
amplitudes contained in $ T^{(1)} $. Indeed, we can write
\begin{equation}
\label{factorised}
A^{(1)}_{pp}A^{(1)}_{\pi\pi}-\left(A^{(1)}_{\pi p}\right)^2=\det
(T^{(1)})=\frac{\det (D)}{\Delta }.
\label{ipso}
\end{equation}
Multiplying both sides of (\ref{ipso}) by $\Delta_k^2$, and using (\ref{sum}),
this gives
\begin{eqnarray}
\lefteqn{\left\{
\left[
a_{pp;\, k}+ \Delta_k\sum_{m\neq k} \left({a_{pp;\, m}\over \Delta_m}\right)
+\Delta_k F_{pp}
\right]
\left[a_{\pi\pi;\, k}+
\Delta_k \sum_{m\neq k} \left({a_{\pi\pi;\, m}\over \Delta_m}\right)
+\Delta_k F_{\pi\pi}
\right]
\right.
\nonumber}\\
\lefteqn{-\left.
\left[a_{\pi p;\, k}+
 \Delta_k\sum_{m\neq k} \left({a_{\pi p;\, m}\over \Delta_m}\right)
+\Delta_k F_{\pi p}\right]
\left[
a_{\pi p;\, k}+\Delta_k\sum_{m\neq k} \left({a_{\pi p;\, m}\over
\Delta_m}\right)
+\Delta_k F_{\pi p}\right]
\right\}\nonumber}\\
&=&
\Delta_k\left\{
{\det(D)\over \Delta_1...\Delta_{k-1}
\Delta_{k+1}...\Delta_N}
\right\}
\end{eqnarray}
In the limit $j\rightarrow z_k$, $\Delta_k\rightarrow 0$, and we get
\begin{equation}
a_{pp;\, k}\ a_{\pi\pi;\, k}-\left(a_{\pi p;\, k}\right)^2=0\quad {\rm for}\
k=1,...,N
\label{facto}\end{equation}
or equivalently
\begin{equation}
\label{factoris}
\lim _{j\rightarrow z_{k}}\left[
A^{(1)}_{pp}(j)-\frac{\left(A^{(1)}_{\pi
p}(j)\right)^2}{A^{(1)}_{\pi\pi}(j)}\right] =\rm finite\: terms.
\end{equation}

The relations (\ref{solut}) and (\ref{factorised}) hold
for all values of $t$ above
the $4 m_p^2$ threshold. Hence they must hold for all $t$ as the amplitudes
are analytic, and in particular they can be continued to negative $t$. The
consequence (\ref{facto}) must thus remain true for the amplitude in the
direct channel.

We show in Appendix 1 that this relation remains the same in
the case of an arbitrary number of elastic and inelastic thresholds:
\begin{equation}
T^{(1)}(\1-RD)=D
\label{solved}
\end{equation}
where $T^{(1)}$ is the matrix containing all elastic amplitudes, and $D$ is
now an unknown matrix containing the effects of inelastic thresholds and
the contribution of elastic amplitudes across the cuts. Again, the
singularities of $T^{(1)}$ must come from the zeroes $z_m$ of
\begin{equation}
\Delta=\det(\1-RD)
\end{equation}
and one obtains again (\ref{facto}) and (\ref{factoris}).
\subsection{Properties}
\begin{enumerate}
\item The factorisation relations are in general broken when one calculates
the contribution of multiple exchanges of trajectories.
However, as the above argument is totally general,  after all
the ($s$-channel) unitarising exchanges are taken into account,
one must end up again with an amplitude factorising at each singularity,
even if the latter is not a simple pole.
\item The matrix $D$ is sensitive to the existence of thresholds
associated with bound states, and does not know directly about quarks and
gluons
which do not enter the unitarity equations. Hence \emph{ the zeroes $z_m$
are not calculable perturbatively.}
\item One could have a spurious cancellation of the
singularity if some element of $ D $ has a zero at $ j=z_{m} $. However,
it is unlikely for this cancellation to occur for
all $ t $ or for all processes. It is however possible to ``hide''
a singularity, \emph{e.g.} at $t=0$ for $ pp $ and $ \bar{p}p $
scattering. This might provide an explanation for the absence of
an odderon pole in forward scattering data.
\item \emph{Each singularity factorises
separately.} Hence it does not make sense to consider globally factorising
cross sections or amplitudes in the $ s $, $ t $ representation, unless
of course the amplitude can be reproduced by only one leading singularity.
\item The relations (\ref{factoris}) lead to a definite prediction
for the residues (or couplings) of the singularities above threshold
$ t>4m_{a}^{2} $. As no singularity occurs when $ t $ is continued
to the physical region for the $ s $ channel processes, these relations
remain true there.
\item We have mentioned that one always obtains a relation between the
amplitude
and its complex conjugate. In fact, this is derived in Appendix 1 as a
consequence of the unitarity of the $S$ matrix. The relations between
amplitudes across a cut are really derived from the relation between
the amplitude and its complex conjugate, and hold for whatever structure
the cuts have.
\end{enumerate}
 \subsection{Specific examples}
\label{spec}
Eq. (\ref{factoris}) is usually not mentioned, and only its
consequences for
the residues of simple poles are considered. However, we have shown
that it is true in general, and that leads to specific predictions
for any singularity structure of the amplitudes $ A^{(1)}_{il}(j) $,
\emph{e.g.} for
a given order of the zeroes $ z_{m} $. We shall give here the
formulae that correspond to simple, double or triple poles, which
seem to be three possibilities emerging from fits to hadronic amplitudes
at $ t=0 $ \cite{compete}. We shall refer to these relations as the $t$-Channel
Unitarity  relations.
The case of cuts will not be explicitly
considered here, although Eq. (\ref{factoris}) holds also in this
case.

For isolated simple poles\begin{equation}
A^{(1)}_{il}=\sum _{m}\frac{R_{il;\, m}}{j-z_{m}},\end{equation}
one obtains the usual relations for the residues
\cite{factorisation}\begin{equation}
R_{22;\, m}=\frac{\left( R_{12;\, m}\right) ^{2}}{R_{11;\, m}}.\end{equation}

If $ A^{(1)}_{il} $ has coinciding simple and double poles\begin{equation}
A^{(1)}_{il}=\frac{S_{il}}{j-z}+\frac{D_{il}}{(j-z)^{2}},\end{equation}
one obtains the new relations
\begin{eqnarray}
D_{11}D_{22}&=&{\left( D_{12}\right) ^{2}},\nonumber\\
\label{double}
D_{11}S_{22}+S_{11}D_{22}&=&{2 D_{12}S_{12}}.
\end{eqnarray}

In the case of triple poles
\begin{equation}
A^{(1)}_{il}=\frac{S_{il}}{j-z}+\frac{D_{il}}{(j-z)^{2}}
+\frac{F_{il}}{(j-z)^{3}},\end{equation}
the relations become
\begin{eqnarray}
F_{11}F_{22}&=&{\left( F_{12}\right) ^{2}},\nonumber\\
\label{triple}
F_{11}D_{22}+D_{11}F_{22}&=&{2 F_{12}D_{12}},\\
D_{11}D_{22}+S_{11}F_{22}+S_{22}F_{11}&=&2S_{12} F_{12} + D_{12}^2.
\nonumber
\end{eqnarray}
It is worth pointing out that the double-pole relations are not the
limit of the triple-pole relations for a vanishing triple-pole residue.
Similarly, the simple-pole relations cannot be obtained from the double-pole
ones. The reason for this is that the relations (\ref{factorised})
relate the poles of order $ 2n $ to $ n+1 $, $n$ being the maximal
order of the pole. Hence the first double-pole relation is contained in
the triple-pole ones, but not the second one, and the simple-pole relations
are entirely separate.
\section{The photon case}
\label{photon}
Photons can also be considered within the formalism of $t$-channel unitarity.
However, it is unclear whether the elastic cross section can be measured, or
even defined \cite{BN}. In practise, what is measured is the hadronic part of
the cross section. Hence only the hadronic part of the photon wave function
enters the measurement, and this part is not directly related to the $S$
matrix.

The way to circumvent this problem is to consider the photons as external
state insertions on the hadronic $S$ matrix. This means that they will
enter the unitarity equations only as external states, and will not
contribute to the thresholds.

In practise, the equations
(\ref{factorised}, \ref{solved}) remain the same, provided that we write
the threshold matrix $ R $ as
\begin{equation}
R=\left( \begin{array}{cc}
\rho _{p} & 0\\
0 & 0
\end{array}\right) .\end{equation}

This means that $ \Delta=1-\rho_p D_{pp} $ only involves $ D_{pp} $, hence
singularities can now come from other elements of $ D $, and $ \det (D) $
can contain singularities not present in $ \Delta , $ hence breaking
the factorisation relations (\ref{ipso},\ref{factoris}). Namely, we obtain
\begin{eqnarray}
A^{(1)}_{pp}&=&\frac{D_{pp}}{\Delta },\nonumber\\
\label{pessimist}
A^{(1)}_{\gamma p}&=&\frac{D_{\gamma p}}{\Delta },\\
A^{(1)}_{\gamma \gamma }&=&\frac{\rho _{p}D_{\gamma p}^{2}}{\Delta }+
D_{\gamma \gamma }.\nonumber
\end{eqnarray}

Extra singularities can come from $ D_{\gamma p} $ or $ D_{\gamma \gamma } $.
In the first case, the nature of the singularity is different in $ \gamma p $
and in $ \gamma \gamma  $, and the coupling of the singularity,
which contains $ \Delta  $, must be of non-perturbative origin.
On the other hand, singularities in $ D_{\gamma \gamma } $ can
be purely perturbative.

In the DIS case, the off-shell photons can also be treated as external
particles,
and one recovers the above equations (\ref{pessimist}) and the
possibility of extra singularities. Details and general formulae are given
in Appendix 2, where one obtains an equation similar to (\ref{pessimist}),
with a matrix $D$ depending on the off-shellnesses of photons:
\begin{equation}
\label{Master2}
T^{(1)}(Q_{in},Q_{out})=D(Q_{in},Q_{out})
+\frac{D(Q_{in},0)RD(0,Q_{out})}{\1-RD(0,0)}.
\end{equation}
where $ Q_{in} $
stands for the two virtualities $ (Q_1^{2},Q_2^{2}) $ of the
initial states in the \break $ t $ channel, and $ Q_{out} $ for the
two virtualities $ (Q_{3}^{2},Q_{4}^{2}) $ of the final states.

We want to point out that the position of the possible new singularities
can depend on $ Q^{2} $, and as the off-shell states do not enter
unitarity equations, these singularities can be fixed in $ t $.
\section{Test of $t$-Channel
Unitarity relations}
\label{testtcu}
In order to test the previous equations, and to evaluate the need
for new singularities, we shall use models that reproduce $ pp $, $ \gamma p $
and $ \gamma \gamma  $ cross sections. Previous studies \cite{compete}
have shown that there are at least three broad classes of models that
can reproduce all forward hadron and photon data.

The general form of these parametrisations is given, for total cross
sections of $a$ on $b$, by the generic formula\footnote{The real part of
the amplitudes, when needed to fit the $\rho$ parameter, is obtained from
derivative dispersion relations \cite{DDR}.}
\begin{equation} \sigma^{tot}_{ab}=(R_{ab}+H_{ab})/s \end{equation}where
$R_{ab}$ is the contribution of the highest meson trajectories ($\rho$,
$\omega$, $a$ and $f$)
and the rising term $H_{ab}$ stands for the pomeron. The first term
is parametrised via Regge theory, and we allow the lower trajectories
to be partially non-degenerate, {\it i.e.} we allow one intercept
for the charge-even trajectories, and another one for the charge-odd ones
 \cite{CKK}.
Hence we use
\begin{equation} R_{ab}= Y_{ab}^{+} \left({\tilde s}\right)^{\alpha _{+}} \pm
Y_{ab}^{-} \left({\tilde s}\right)^{\alpha _{-}}\label{lower} \end{equation}
with
$\tilde s=2\nu/(1$ GeV$^2)$, $\nu=p.q$.

As for the pomeron term, we consider the following possibilities:
\begin{eqnarray}
H_{ab}&=&X_{ab}\left[ \widetilde{s}\right] ^{\alpha _{\wp }},\\
\label{doubpar}
H_{ab}&=&{\tilde s}D_{ab}\left[ \log \widetilde{s}+\log C_{ab}\right], \\
\label{trippar}
H_{ab}&=&{\tilde s}t_{ab}\left[ \log^2
\left(\frac{\widetilde{s}}{d_{ab}}\right)
+\log \left( c_{ab}\right) \right] .
\end{eqnarray}
These forms come from simple, double or triple poles in the
Watson-Som\-mer\-feld transform of the amplitude (see Eq. (\ref{Regge}) of
Appendix 1),
in the limit of $ \cos (\vartheta _{t}) $ large, so that the contribution
from the integration contour vanishes, and that one can keep only
the leading meson trajectories and the pomeron contribution.

Using the asymptotic expansion of the Legendre polynomials $ P_{l} $
\begin{equation}
\label{Legendre}
P_{l}(-\cos (\vartheta _{t}))\rightarrow \frac{\Gamma (2l+1)}{[\Gamma
(l+1)]^{2}2^{l}}\left( \frac{\nu }{m_{p}^{2}}\right) ^{l},
\end{equation}
 we obtain, by the residue theorem (see Eq. (\ref{Regge}) of Appendix 1),
the following
contributions to the total cross section for simple, double, and triple
poles:
\begin{eqnarray}
\label{SWsimp}
A^{(1)}(j,0)&=&\frac{g}{j-\alpha }\nonumber\\
&\rightarrow& \sigma_{tot}=g\left( \frac{\nu
}{m_{p}^{2}}\right) ^{\alpha }\frac{(2\alpha +1)\Gamma (2\alpha +1)}{(\Gamma
(\alpha +1))^{2}2^{\alpha }},\\
\label{SWdoub}
A^{(1)}(j,0)&=&\frac{g}{(j-\alpha )^{2}}\nonumber\\
&\rightarrow& \sigma _{tot}=g\left( \frac{\nu
}{m_{p}^{2}}\right) ^{\alpha }\log \left( \frac{\nu }{m_{p}^{2}}\right)
\frac{(2\alpha +1)\Gamma (2\alpha +1)}{(\Gamma (\alpha +1))^{2}2^{\alpha }},\\
\label{SWtrip}
A^{(1)}(j,0)&=&\frac{g}{(j-\alpha )^{3}}\nonumber\\
&\rightarrow& \sigma _{tot}=g\left( \frac{\nu
}{m_{p}^{2}}\right) ^{\alpha }\log^2 \left( \frac{\nu }{m_{p}^{2}}\right)
\frac{(2\alpha +1)\Gamma (2\alpha +1)}{(\Gamma (\alpha +1))^{2}2^{\alpha
+1}}.
\end{eqnarray}

In the photon case, things are a little different. Looking first at
the $ \gamma p $ amplitude with off-shell photons, we have \begin{equation}
\label{cosg}
|\cos (\vartheta _{t})|=\frac{\nu }{m_{p}\sqrt{Q^{2}}}.
\end{equation}

In the on-shell limit $ Q^{2}\rightarrow 0 $, the Legendre polynomial
of Eq. (\ref{Legendre}) becomes infinite, hence one must assume that
the amplitude goes to zero in such a way that the limit is finite.
One can take for instance\begin{equation}
\label{aigpl}
A^{(1)}_{\gamma p}=\tilde{A}^{(1)}_{\gamma p}\left(
\frac{\sqrt{Q^{2}}}{q_{\gamma
}(Q^{2})}\right) ^{j}
\end{equation}
with $ q_{\gamma }(0) $ finite. Such a choice introduces a new
scale that effectively replaces $ \sqrt{Q^{2}} $
with $ q_{\gamma }(Q^{2}) $ in $ \cos (\vartheta _{t}) $, and
$ A^{(1)} $ with $ \tilde{A}^{(1)} $. In
the $ \gamma \gamma  $ case, with $Q^2$ and $P^2$ the off-shellnesses
of the two incoming photons, in order to keep the unitarity
relations (\ref{factorised})
for the amplitude $ \tilde{A}^{(1)} $ instead of $ A^{(1)} $, one needs to
assume that \begin{equation}
\label{aigpl2}
A^{(1)}_{\gamma \gamma }=\tilde{A}^{(1)}_{\gamma \gamma }\left(
\frac{\sqrt{Q^{2}P^2}}{q_{\gamma }(Q^{2})q_{\gamma }(P^{2})}\right) ^{j}
\end{equation}
and the scales $q_{\gamma }(Q^{2}) $  and $ q_{\gamma }(P^2)$ replace $ m_{p}$
in Eqs. (\ref{SWsimp}-\ref{SWtrip}).
\subsection{Regge region}
\label{reggereg}
One can think of translating the minimum $ \sqrt{s} $ of
the $ pp $ case into a bound for $ {\nu }/{m_{p}^{2}} $,
$ {\nu }/({m_{p}q_{\gamma }(Q^{2})}) $ and $ {\nu }/{q_{\gamma }^{2}(Q^{2})} $,
and use the same bound in the three processes. Unfortunately, the
situation is really more complicated because one cannot extract $q_\gamma(Q^2)$
from the data as the
$\log\nu$ terms come from  a combination of simple, double (and triple) poles
at $j=1$, which can always be re-shuffled among themselves.

In the following, we shall use cuts on the natural Regge
variables $2\nu$, and
$\cos(\vartheta_t)$. We find that data are well reproduced in the region
\begin{eqnarray}
\cos(\vartheta_t) & \ge & \frac{49 {\rm\ GeV}^2}{2m_p^2},\\
\sqrt{2\nu}  & \ge & 7\ \rm GeV.
\end{eqnarray}
For the $\gamma \gamma$ and the $\gamma p$ total
cross sections, as well as for the photon structure function
where $P^2 \rightarrow 0$,
$\cos(\vartheta_t)$ becomes infinite, and only the cut on $2\nu$ constrains
the Regge region.

Furthermore, in the case of one virtual photon, experimentalists measure
the $ ep $ or the $ e\gamma  $ cross sections. From these, one
can extract a cross section for $ \gamma ^{*}p $ or $ \gamma ^{*}\gamma ^{*} $
scattering, provided one factors out a flux factor. As is well known,
the latter is univoquely defined only for on-shell particles:\begin{equation}
\label{sigf2}
\sigma_{tot} =\lim_{Q^2\rightarrow 0}\frac{4\pi ^{2}\alpha }{Q^{2}}F_{2}.
\end{equation}
The flux factor can then be modified arbitrarily, provided that the
modifications vanish as $ Q^{2}\rightarrow 0 $. This means, for
instance, that we can always multiply the left-hand side of (\ref{sigf2})
by an arbitrary power of $ (1-x) $. Hence one should in principle
limit oneself to small values of $ x $ only.
We find that we can obtain good fits in the region
\begin{equation}
x\leq 0.3.
\end{equation}
Note that in the case of two
off-shell photons, experimentalists measure $ \sigma _{TT}+\sigma _{TL}+\sigma
_{LT}+\sigma _{LL} $,
which is precisely the quantity entering the factorisation theorem.
Hence no flux factor is necessary here.

Finally, all the residues are expected to be functions of $Q^2$. These form
factors are unknown, and are expected to contain higher twists.
In order to check factorisation, we do not want to be too dependent on these
guesses. Hence we choose a modest region of
\begin{equation}
Q^2\leq 150 \ {\rm GeV}^2,
\end{equation}
where most of the $\gamma^*\gamma^*$ points lie.

We shall consider in the next section possible extensions to a wider region.
\subsection{Factorising $ t $-Channel
Unitarity relations }
\label{factorising}
As explained above, the simple-pole singularities will factorise in the
usual way. Note that there is no charge-odd singularity in the photon
case, hence only the $ a/f $ lower trajectory will enter the relations.
One then gets \begin{equation}
\label{lowerf}
Y_{pp}Y_{\gamma \gamma }(P^{2},Q^{2})=Y_{\gamma p}(P^{2})Y_{\gamma p}(Q^{2}).
\end{equation}

In the case of a soft-pomeron pole, one obtains similarly\begin{equation}
\label{softpom}
X_{pp}X_{\gamma \gamma }(P^{2},Q^{2})=X_{\gamma p}(P^{2})X_{\gamma p}(Q^{2}).
\end{equation}

The case of multiple poles is given by Eqs. (\ref{double}, \ref{triple}),
and can be made more transparent by using the forms (\ref{doubpar},
\ref{trippar}) which give factorisation-looking relations for the
constants (but not for all the residues! $-$ see Eqs. (\ref{double})
and (\ref{triple})~$-$):
\begin{equation}
f_{pp}f_{\gamma \gamma }(P^{2},Q^{2})=f_{\gamma p}(P^{2})f_{\gamma
p}(Q^{2})\end{equation}
with $f=D$, $C$, $t$, $d$ or $c$.
\subsection{Dataset}
\label{dataset}
For the total cross sections, we have used the updated COMPETE dataset used in
\cite{PRL}, which is the same as that of \cite{RPP} except for
the inclusion of the latest ZEUS results on $\gamma p$ cross section
\cite{ZEUS}
and for the inclusion of cosmic-ray data.

For $\gamma p$ scattering, we have used the full set of available data
\cite{H1,ZEUS,others}.

For the $\gamma\gamma$ measurements of $F_2^\gamma$, we have used the data
of \cite{L3,OPAL}, whenever these included
the joint $x$ and $Q^2$ (and $P^2$) dependence. We have not included other
data as they do not have points in the Regge region.
Note that we have not taken the
uncertainties in $x$ into account, hence the $\chi^2$ values are really
upper bounds in the $\gamma\gamma$ case.
\subsection{Previous parametrisations}
\label{previous}
We have first considered the results using previous studies \cite{DM,CS}
of $ \gamma ^{(*)}p $ and $ pp $ scattering. Making use of the $t$-Channel
Unitarity
relations (\ref{double}) and (\ref{triple}), we have obtained
reasonably good predictions for $ \sigma _{\gamma \gamma } $ and
$ F_{2}^{\gamma } $. However, the formalism breaks
down in the case of $ \gamma ^{*}\gamma ^{*} $ scattering, because
the form factors that we used do not guarantee the positivity of the
charge-even part of the
cross sections. Re-fitting them enables one to get closer to the data,
but the problem of negativity remains in some part of the physical
region. Hence, at this point, the factorisation relations have one major
consequence: the parametrisations of \cite{DM,CS} are ruled out.

We have also considered the hard pomeron fit of \cite{DoLa}
where the charge-parity $+1$
rising term contains two different simple poles: the soft and the hard pomeron.
In this case, the soft pomeron residues factorise. The hard pomeron, if
it is
not present in $pp$ cross sections (see however \cite{clms}),
then comes in as a double
pole in $\gamma\gamma$ cross sections, see Eq. (\ref{pessimist}), and
produces a cross section proportional to $\nu^{\alpha_h}\log\nu$.
Its residue will then depend on the value of $\Delta(j=\alpha_h)$,
which is unknown. This means that factorisation does not say much about
the hard pomeron contribution, which can always be arbitrarily re-scaled.
It is possible to get good fits using these forms, but as they do not
test factorisation, we shall not present these results here.
\subsection{New parametrisation: triple pole}
\label{new}
\begin{figure}
\centering{\epsfig{file=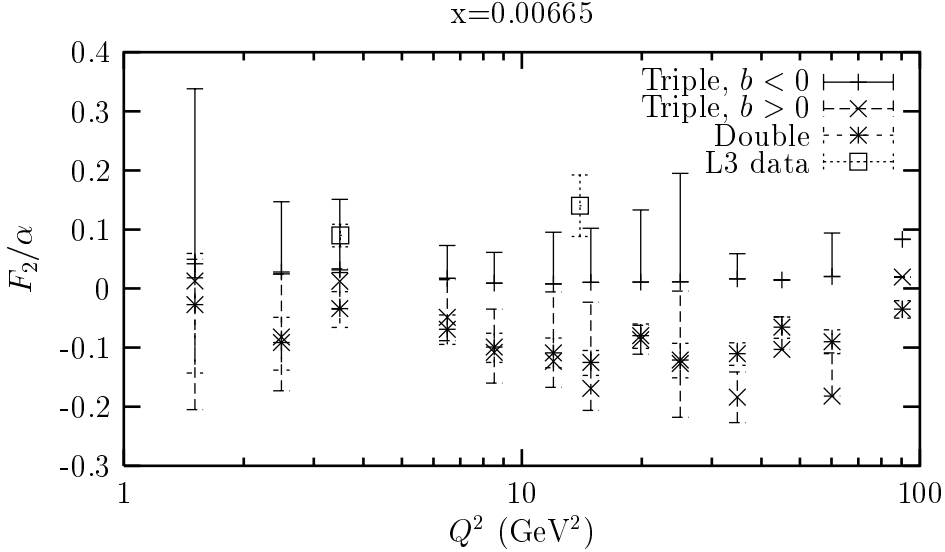}}
\caption{Prediction from $t$-Channel
Unitarity relations for the $\gamma^*\gamma^*$ cross
section, including the box diagram of Appendix 3.}
\label{2branch}
\end{figure}

In the triple-pole case, the problem of negativity
can be cured through the introduction
of another functional form for the form factors. To convince ourselves
that this is possible, we have fitted $ F_{2} $ in several $ Q^{2} $
bins to\begin{equation}
F_{2}^{p}(Q^{2})=a(\log \nu +b)^{2}+c\nu ^{-0.47}.\end{equation}
{}From the values of $ a, $ $ b $ and $ c $, and the $t$-Channel
Unitarity relations,
one can then predict
the symmetric $ F_{2}^{\gamma }(Q^{2},Q^{2}) $. The result
of this exercise is shown in Fig. \ref{2branch}.
One clearly sees that there are two branches in the fit to HERA data:
one with positive $ b $, and another one with negative $ b $.
Both have comparable $ \chi ^{2} $, but one produces positive $ \gamma \gamma
$
cross sections, whereas the other one does not. Armed with this information,
we found that the resulting form factors could be well approximated
by the following forms: \begin{eqnarray}
t_{\gamma p}(Q^{2})&=&t_{1}\left( \frac{1}{1+\frac{Q^{2}}{Q_{t}^{2}}}\right)
^{\epsilon _{t}},\nonumber\\
\label{formfact}
Y^{+}_{\gamma p}(Q^{2})&=&Y_{1}\left(
\frac{1}{1+\frac{Q^{2}}{Q_{y}^{2}}}\right) ^{\epsilon _{y}},\\
\log d_{\gamma p}(Q^{2})&=&d_{1}\left( \frac{Q^{2}}{Q^{2}+Q_{d}^{2}}\right)
^{\epsilon _{d}},\nonumber\\
\log c_{\gamma p}(Q^{2})&=&c_{0}+c_{1}\left(
\frac{1}{1+\frac{Q^{2}}{Q_{c}^{2}}}\right) ^{\epsilon
_{c}}.\nonumber\end{eqnarray}
With the form factors obtained from our fit, we have then checked
that the $ \gamma ^{*}\gamma ^{*} $ cross section remains positive
everywhere.
\subsection{New parametrisation: double pole}
\label{newdp}
In the case of a double pole, Fig. \ref{2branch} shows that the situation
is more difficult, as one cannot guarantee positivity. We have tried
several possibilities, among which a further splitting of leading
meson trajectories along the lines of \cite{DM2}, but found that
positivity is still not guaranteed.

However, it is possible to obtain a good fit, positive everywhere,
if one assumes a slightly modified version
of the double pole \cite{DM3}.

Instead of taking an $\tilde s D\log \tilde s  $ term in $H_{ab}$ as
in Eq.~(\ref{doubpar}),
 one can consider
\begin{equation}
H_{ab}=\tilde s D_{ab}\ [L_{ab} +\log C_{ab}]
\end{equation}
with
\begin{equation}  L_{ab}(\tilde
s)=\frac{1}{2}\Re e[\log(1+\Lambda_{ab}\tilde
s\,^{\delta})+\log(1+\Lambda_{ab}(-\tilde s)\,^{\delta})].
\end{equation}
Asymptotically, this gives the same form as a double pole. Furthermore,
one can rewrite
$\log(1+\Lambda_{ab}(\tilde s)\,^{\delta})=\delta \log(\tilde s)+
\log(\Lambda_{ab}+1/(\tilde s)\,^{\delta})$. The first term comes
from a double pole at $j=1$, whereas the Taylor expansion of the remaining
term gives a series of simple poles. Hence $D_{ab}$ and $\Lambda_{ab}$
factorises according to
\begin{eqnarray}
D_{\gamma\gamma}(P^2,Q^2)D_{pp}&=&D_{\gamma p}(P^2) D_{\gamma
p}(Q^2),\nonumber\\
\Lambda_{\gamma\gamma}(P^2,Q^2)\Lambda_{pp}&=&\Lambda_{\gamma p}(P^2)
\Lambda_{\gamma p}(Q^2).
\end{eqnarray}
We found good fits using the following form factors:
\begin{eqnarray}
D_{\gamma p}&=&D_1\left (\frac{1}{1+\frac{Q^{2}}{Q_{d}^{2}}} \right
)^{\epsilon_{d}},\nonumber\\
\label{formfacd}
D_{\gamma p}\log C_{\gamma p}&=&C_1\left
(\frac{1}{1+\frac{Q^{2}}{Q_{c}^{2}}} \right )^{\epsilon_{c}},\\
\Lambda_{\gamma p}&=&\Lambda_1\left
(\frac{1}{1+\frac{Q^{2}}{Q_{\lambda}^{2}}} \right
)^{\epsilon_{\lambda}},\nonumber\\
Y^{+}_{\gamma p}(Q^{2})&=&Y_{1}\left(
\frac{1}{1+\frac{Q^{2}}{Q_{y}^{2}}}\right) ^{\epsilon _{y}}.\nonumber
\end{eqnarray}
\begin{table}[ht]
\small
\centering{\begin{tabular}{|l|c|c|c|c|c|}
\hline
 & & \multicolumn{2}{c|}{double} & \multicolumn{2}{c|}{triple} \\
\hline
Quantity & $N$ & $\chi^2$ & $\chi^2/N$ & $\chi^2$ & $\chi^2/N$ \\
\hline
$F_2^p$        &  821 &  789.624 & 0.962 &  870.599 & 1.060 \\
$F_2^\gamma$        &   65 &   57.686 & 0.887 &   59.963 & 0.923 \\
$\sigma_{\gamma\gamma}$  &   32 &   19.325 & 0.604 &   15.568 & 0.487 \\
$\sigma_{\gamma p}$  &   30 &   17.546 & 0.585 &   21.560 & 0.719 \\
$\sigma_{pp}$  &   90 &  100.373 & 1.115 &   82.849 & 0.921 \\
$\sigma_{p\bar p}$ &   49 &   55.240 & 1.127 &   58.900 & 1.202 \\
$\rho_{pp}$    &   67 &   93.948 & 1.402 &   98.545 & 1.471 \\
$\rho_{p\bar p}$   &   11 &   16.758 & 1.523 &    4.662 & 0.424 \\
\hline
Total          & 1165 & 1150.500 & 0.988 & 1212.645 & 1.041 \\
\hline
\end{tabular}}
\caption{Results of fits to a generalised double pole model and to a triple
pole model, using the form factors of Eqs.~(\ref{formfact}) and
(\ref{formfacd}). $N$ is the number of data points for each sub-process.}
\label{tab1}
\end{table}
\subsection{Box diagram}
\label{boxdi}
\begin{figure}
\centering{\epsfig{file=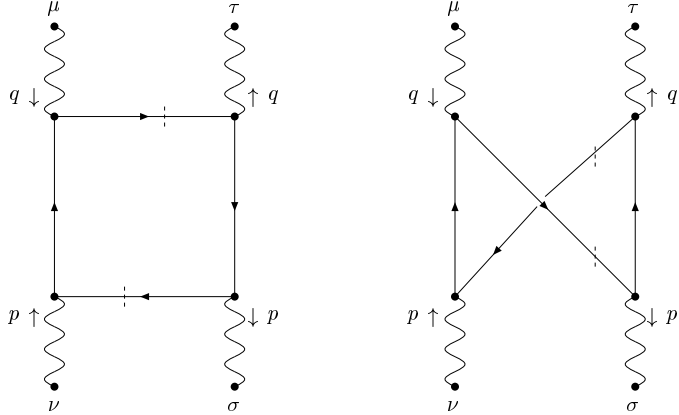,width=9cm}}
\caption{The box diagram contribution.}
\label{box}
\end{figure}
One new singularity may be present in $\gamma\gamma$ scattering:
it is the box diagram,
shown in Fig. \ref{box}, which couples directly two photons to quarks.
This diagram must be present when the photons are far
off-shell and
pQCD applies. As we have explained above, it is not at all obvious that
it is present in the case of total cross sections, and
in fact we get better fits if we include it only for off-shell photons.
Hence an extra perturbative singularity is needed at nonzero $Q^2$.

We have re-calculated it and confirm the results of \cite{Budnev} (see
Appendix 3\footnote{
 We want
to point out that one need to calculate $\sigma_{LL}$, $\sigma_{TL}$,
$\sigma_{LT}$
and $\sigma_{TT}$ separately and sum them to obtain the off-shell cross
section. A contraction of $g_{\mu\nu}$ does not re-sum the helicities
properly \cite{peskin} which probably explains the discrepancies between
\cite{Budnev} and \cite{DoDo}.}).
In the following, we shall fix the quark masses at
\begin{eqnarray}
m_u=m_d&=&0.3 {\ \rm GeV},\nonumber\\
m_s&=&0.5 {\ \rm GeV},\\
m_c&=&1.5 {\ \rm GeV},\nonumber\\
m_b&=&4.5 {\ \rm GeV}.\nonumber
\end{eqnarray}
and the quarks are included only above threshold $s=2\nu-P^2-Q^2>4m_q^2$.
\subsection{Results}
As we want to be able to vary the minimum value of $2\nu$,
and as the fits of \cite{compete} neither include the generalised dipole
nor use $2\nu$ as the energy variable,
we have re-fitted the $pp$ and $\bar pp$
cross sections and $\rho$ parameter together with those for $\gamma^{(*)} p$
and $\gamma^{(*)}\gamma^{(*)}$, and imposed factorisation of the residues.
\begin{table}[ht]
\centering{\begin{tabular}{|l|c|l|c|}
\hline
\multicolumn{2}{|c|}{triple} & \multicolumn{2}{c|}{double} \\
\hline
Parameter & Value  & Parameter & Value  \\
\hline
\hline
$t_{pp}$           &0.6264 $\pm$0.0055&$\Lambda_{pp}$  &$1.36 \pm 0.15 $ \\
$log(d_{pp})$      &0.534  $\pm$0.044 &$D_{pp}$     & $40.3 \pm 1.4 $ \\
$t_{pp}\log c_{pp}$&65.86  $\pm$0.48  &$D_{pp}\log C_{pp}$&$-32.7\pm 5.3$\\
$Y^+_{pp}$  & 122.0  $\pm$  1.5       &$Y^+_{pp}$      &  $231.1 \pm 4.7 $ \\
$\alpha_+$         &0.6905$\pm$0.0023&$\alpha_+$   & $0.7263 \pm 0.0010$  \\
 $Y^-_{pp}$       &84.6 $\pm$4.1   &$Y^-_{pp}$   & $97.6 \pm 4.6 $ \\
$\alpha_-$        &0.4596$\pm$0.0010&$\alpha_-$   &  $0.505 \pm 0.015$  \\
\cline{1-1} \cline{2-2}
$c_0$ & $-613.93 \pm  0.91$  & $\delta$ &$0.3313 \pm 0.0092$  \\
\cline{3-3} \cline{4-4}
$c_1$ & 740.8 $\pm$1.2      &$C_1$        &  $-0.105 \pm 0.016$  \\
$Q^2_c$  & 0.1557    $\pm$0.0030 & $Q^{2}_{c}$    &   $0.0219 \pm 0.0076$  \\
$\epsilon_c$ & 0.11619   $\pm$ 0.00047 & $\epsilon_c$ &   $0.553 \pm 0.025$  \\
\hline
$t_1$ &0.001667$\pm$0.000011& $\Lambda_1$       &   $1.49     \pm 0.23  $ \\
$Q^2_t$   & 0.964 $\pm$  0.016 &$Q^2_{\lambda}$ &   $0.111    \pm  0.032$  \\
$\epsilon_t$ & 0.8237$\pm$  0.0034 &$\epsilon_{\lambda}$ &  $0.658 \pm 0.019$\\
\hline
$d_1$  & -8.067$\pm$0.033 & $D_1$        &   $0.1305 \pm 0.0062$  \\
$Q^2_d$         & 7.56   $\pm$0.25  & $Q^2_d$    &   $0.379     \pm 0.061$  \\
$\epsilon_d$ & 0.3081 $\pm$  0.0059 &$\epsilon_d$ & $0.434  \pm 0.021$  \\
\hline
$Y_1$  & 0.1961  $\pm$0.0031 & $Y_{1}$   & $0.515 \pm 0.017$  \\
$Q^2_y$ & 2.056  $\pm$  0.067 & $Q^{2}_{y}$  & $0.158  \pm 0.016$  \\
$\epsilon_y$ & 0.5448    $\pm$  0.0049 &$\epsilon_y$ & $0.709  \pm 0.016$  \\
\hline
\end{tabular}\\~\\}
\caption{Parameters (in natural units) of the global fits.}
\label{tab2}
\label{results}
\end{table}
\begin{figure}[ht]
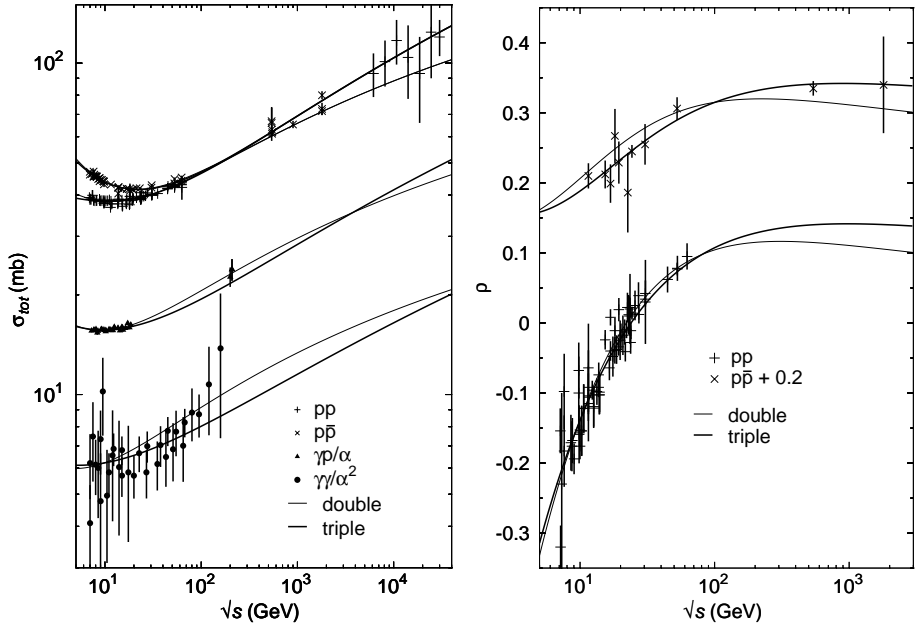

\centering{\epsfig{file=stot.eps,width=6cm}
\epsfig{file=rho.eps,width=6cm}}
\caption{Fits to the total cross sections and to the $\rho$ parameters.}
\label{stot}
\end{figure}
\begin{figure}[ht]
{\epsfig{file=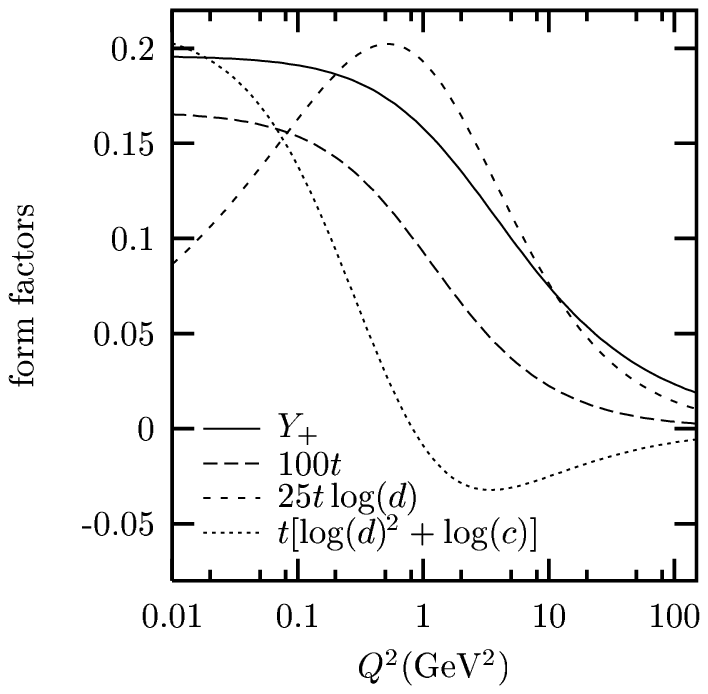,width=6cm}}{\epsfig{file=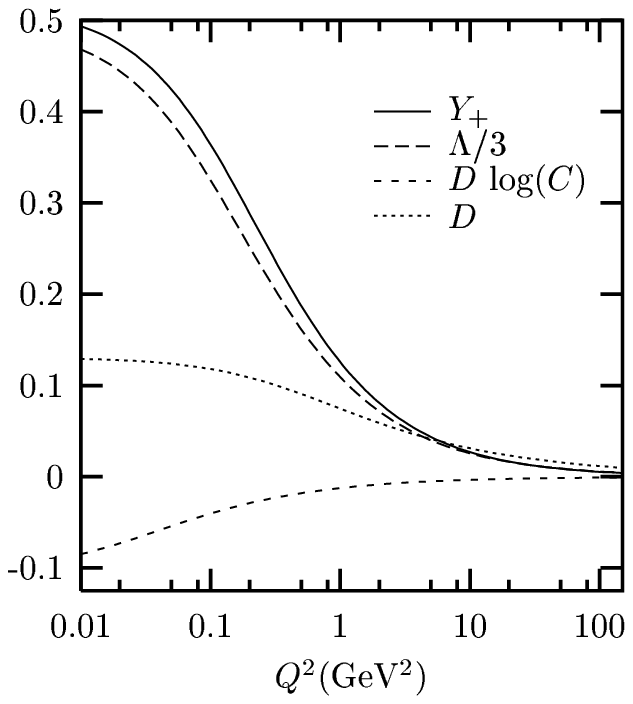,width=6cm}}
\caption{Form factors of the triple pole (left) and double pole (right)
 parametrisations.}
\label{tform}
\end{figure}

We show in Table~\ref{tab1} the $\chi^2$ per data point and
the number of points for each
process.
We see that one obtains a very good global $\chi^2$ for both models. It
is well known \cite{compete} that the partial $\chi^2$ for $\sigma_{p\bar p}$
and $\rho_{pp}$ never reach low values, presumably because of the presence
of contradictory data. We show the corresponding curves in Fig.~\ref{stot}.

The values of the parameters are given in Table \ref{tab2} for the triple-pole
and the double-pole cases, and the
form factors are plotted in Fig. \ref{tform}.

We see that the intercepts of the leading meson trajectories are close,
in fact closer than those of \cite{compete}. This is due to the smaller
energy region, and to the much larger influence of photon data on
$\alpha_+$.

It may also be
noted, in the double-pole case, that the parameter $\delta$ is close to
the hard pomeron intercept of \cite{DoLa}.
At high $Q^2$, because the form
factor $\Lambda$ falls off, the logarithm starts looking like a power of
$2\nu$, and somehow mimics a simple pole. It may thus be thought of as a
unitarized version of the hard pomeron, which would in fact apply to
hard and soft scatterings.

In the triple-pole case, this is accomplished by a different mechanism: the
scale of the logarithm is a rapidly falling function of $Q^2$, and hence the
$\log^2$ term becomes relatively more important at high $Q^2$.
Interestingly, when one writes the triple-pole parametrisation as a function
of $x$ and $Q^2$, one obtains only very small powers (of the order of
0.1) of $Q^2$, which do not contain any higher twists, contrarily to
the soft pomeron of \cite{DoLa}.
\subsection{Total $\gamma p$ and $\gamma\gamma$ cross sections}
\label{total}
We see from Table \ref{tab1} that one obtains an excellent
$ \chi ^{2} $ for $ \sqrt{2\nu }>7 $ GeV, for a total of 62 points.
The curves are shown
in Fig. \ref{stot}. The fit can in fact be continued
to $ \sqrt{2\nu }=2 $ GeV, with a $ \chi ^{2} $/point of 0.74
for 219 points.

We have checked that adding the box diagram leads to a slight degradation of
the
fit, whether one fits the total cross sections alone or with all other data.
As the contribution of the box is calculated perturbatively,
one might object that one cannot use the result down to $ Q^{2}=0 $, and that
only the $\nu$ dependence should be kept.
Hence we have also tried to add an extra term,
proportional to $ \log \nu /\nu  $
in the total cross section, but found that the fit
prefers to set the proportionality constant to zero. Hence it seems
that this singularity is not needed at $P^2=Q^2=0$.
However, because of large uncertainties
in the data, it is not possible to rule it out altogether.

Similarly,
we do not find the need to introduce any new rising contribution.
However, it is clear in view of the large uncertainties that it is
not possible to rule out completely such a possibility. In fact, our
fit prefers the $\gamma\gamma$ data unfolded with PHOJET \cite{PHOJET},
which rise more slowly than those unfolded with PYTHIA \cite{PYTHIA}.
Interestingly, as we reproduce both HERA and LEP data, for $Q^2$
nonzero, it is not true that an extrapolation of the nonzero $Q^2$ data
leads to a higher estimate of the $\gamma p$ and $\gamma\gamma$
cross sections. Our fit
can be considered as an explicit example for which such an
extrapolation leads to a cross section on the lower side of the
experimental errors.
\begin{figure}[ht]
\epsfig{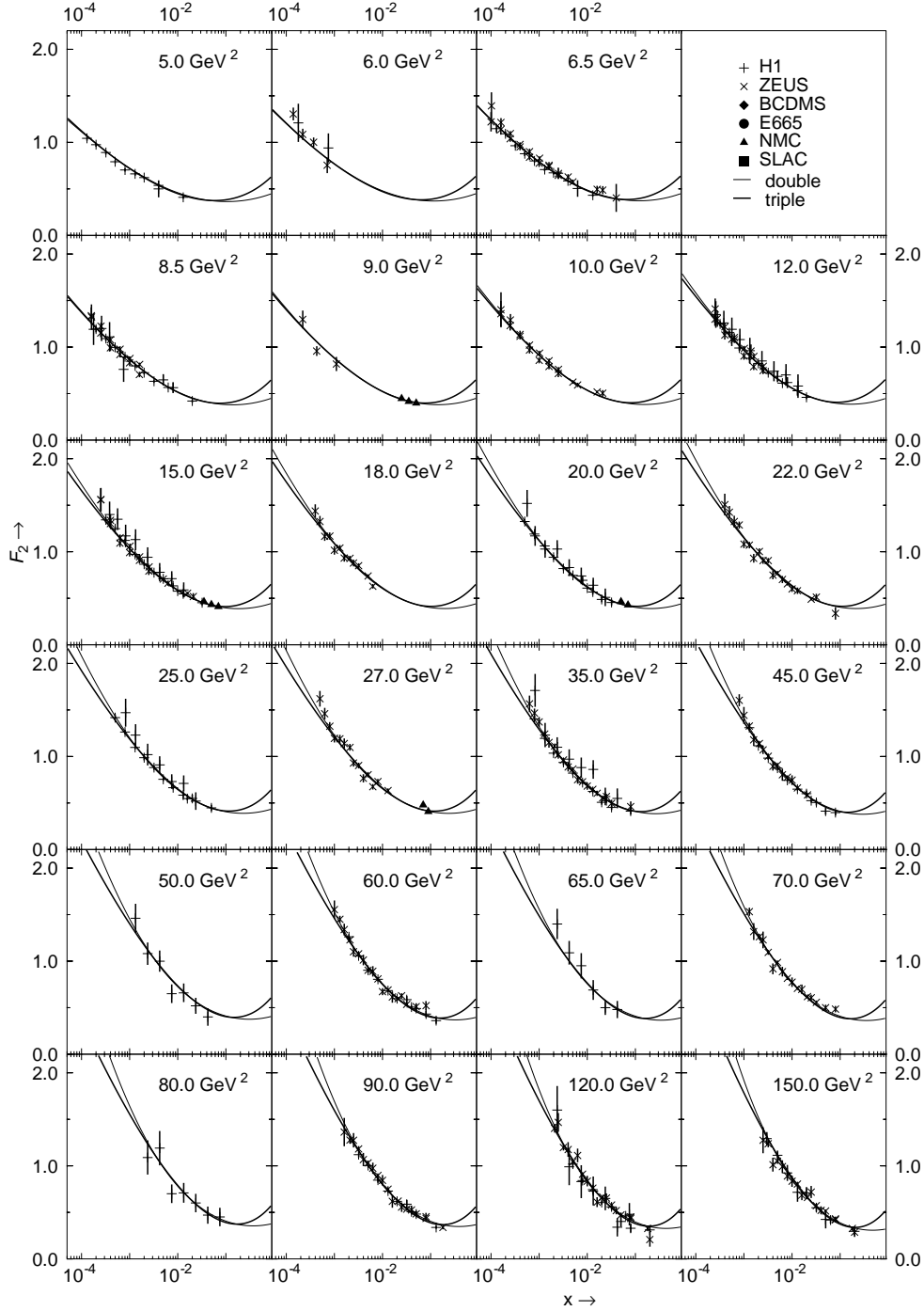}
\caption{Fits to $F_2^p$ in the high-$Q^2$ region. We show only
graphs for which there are more than 6 experimental points.
}
\label{highq2}
\end{figure}
\begin{figure}[ht]
\epsfig{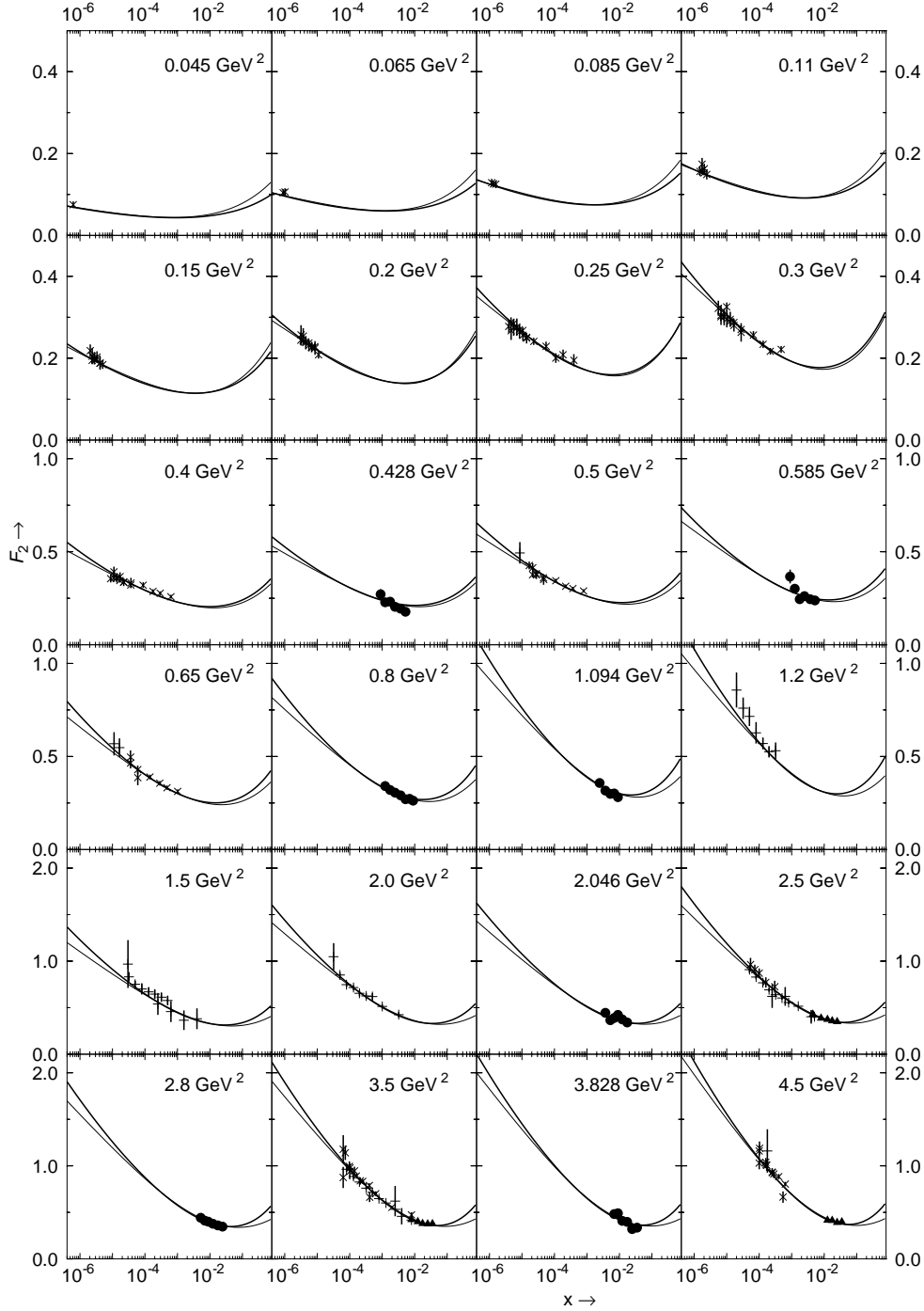}
\caption{Fits to $F_2^p$ in the low-$Q^2$ region. We show only
graphs for which there are more than 6 experimental points, as well as the
lowest-$Q^2$ ones. The curves and the data are as
in Fig.~\ref{highq2}.}
\label{lowq2}
\end{figure}

\vbox{~\vglue 12pt \subsection{$F_2^p$}}
\label{f2p}
The fit to $F_2$ has quite a good $\chi^2$ as well. We have checked that one
can easily extend it to $Q^2\approx 400$ GeV$^2$ for the triple pole, and to
$Q^2\approx 800$ GeV$^2$ in the double-pole case. It is interesting
that one cannot go as high as in ref. \cite{CS}.
This can be attributed either to too simple a choice for
the form factors, or to the onset of perturbative evolution.

Figs. \ref{highq2} and \ref{lowq2} show the $F_2^p$ fit for
the most populated $Q^2$ bins. As pointed out before, we see that
our fits do reproduce the low-$Q^2$ region quite well, but predict
total cross sections on the lower side of the error bands. Hence
the extrapolation to $Q^2=0$ of DIS data does not require a hard pomeron.
\subsection{Fits to $F_2^\gamma$}
\begin{figure}[ht]
\epsfig{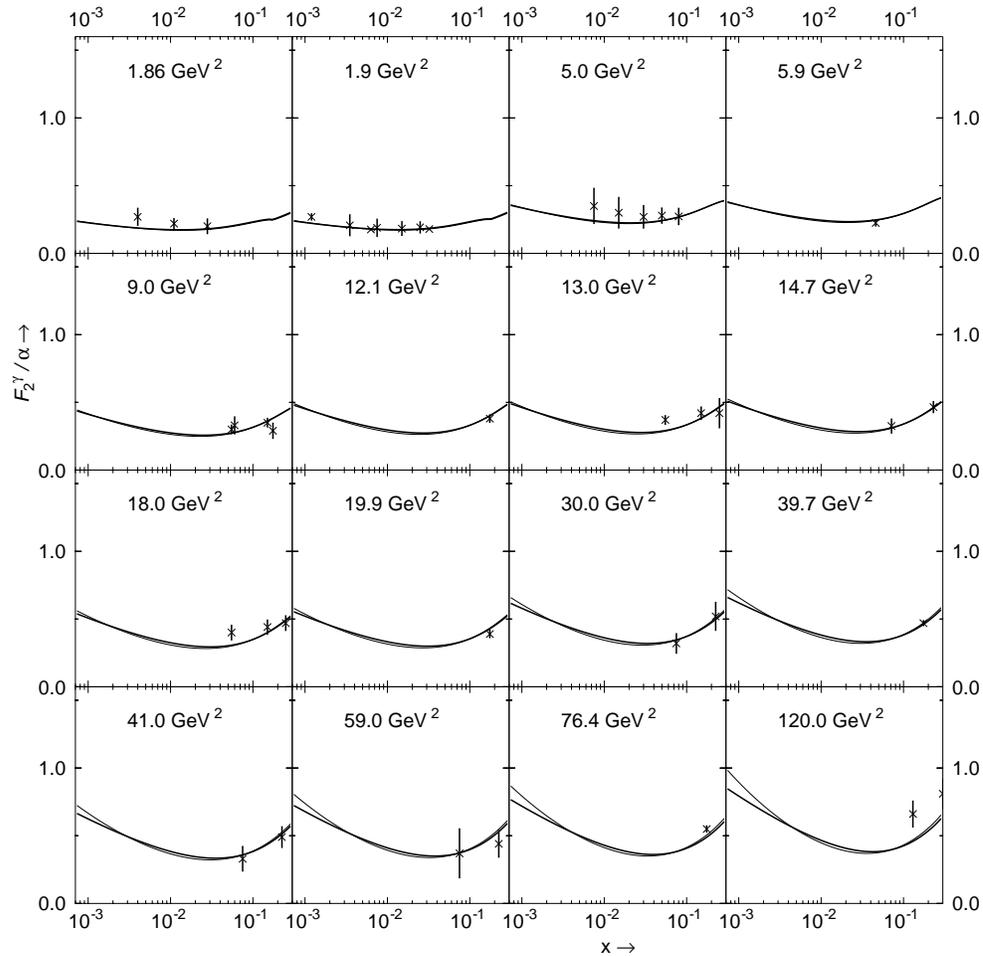}
\caption{Fits to $F_2^\gamma$. The curves are as in Fig.~\ref{highq2}.
The data are from \cite{L3,OPAL}.}
\label{gamma}
\end{figure}
\begin{figure}
\centering{\epsfig{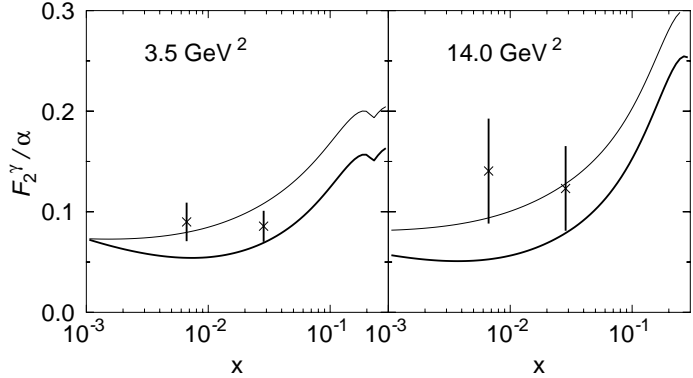}}
\caption{Fits to $F_2^\gamma$ for $P^2=Q^2$. The curves are as in
Fig.~\ref{highq2}.
The data are from \cite{L3}.}
\label{gsame}
\end{figure}
\begin{figure}
\centering{\epsfig{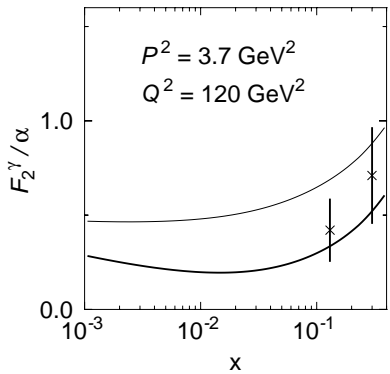}}
\caption{Fits to $F_2^\gamma$ for nonzero asymmetric values of $P^2$ and
$Q^2$. The curves are as in Fig.~\ref{stot}.
The data are from \cite{L3}.}
\label{g37120}
\end{figure}
\label{f2g}
As the number of data points is dominated by $pp$ and $\gamma p$ data, the
fit to $\gamma\gamma$ data is really a test of the $t$-Channel
Unitarity relations. As we
explained
above, the strongest constraint comes from the positivity of the $\gamma^*
\gamma^*$ cross section, which is not guaranteed by the $t$-Channel
Unitarity relations in the
case
of multiple poles. As Fig. \ref{gamma}, \ref{gsame} and \ref{g37120} show, one
obtains a good description of the points within the Regge region.

Here, we have observed that the quality of the fit improves if
we add the box diagram for nonzero $Q^2$ and $P^2$.
There is no need however to include
other singularities, such as a hard pomeron or a perturbative one.

For $Q^2\neq 0$ and $P^2=0$, the box diagram makes little difference in
the double-pole case, but does reduce the $\chi^2$ appreciably in the
triple-pole case. We have included it in both cases.
\section{Conclusion}
\label{conclu}
We have shown in this paper that $t$-channel unitarity can be used to
map the regions where new singularities, be they of perturbative
or non-perturbative origin, can occur. Indeed, we have seen that
although hadronic singularities must be universal, this is certainly
not the case in $\sigma_{\gamma p\rightarrow hadrons}$, $F_2^p$
or $F_2^\gamma$, as the photons enter only as external particles
through an insertion
in the hadronic cross section.

We have shown however that up to\footnote{
The region we have considered excludes the highest-$Q^2$ $\gamma\gamma^*$
points from OPAL. For the point which falls in the Regge region,
at $P^2=0$, $Q^2=780$ GeV$^2$ and $x=0.275$,
the experimental value is $0.93\pm 0.16$,
the extrapolation of the double-pole fit predicts $0.71$,
while the triple-pole prediction is $0.74$.
} $Q^2=150$ GeV$^2$, the data do not call for the existence of new
singularities, except perhaps
the box diagram.

For
off-shell photons, our fits are rather surprising as the standard claim is that
the perturbative evolution sets in quite early.
This evolution is indeed allowed by $t$-channel unitarity constraints: it is
possible to have extra singularities in off-shell photon cross sections, which
are built on top of the non-perturbative singularities. But it seems that Regge
parametrisations can be extended quite
high in $Q^2$ without the need for these new singularities.

Finally, as the BFKL singularity is purely perturbative (the position of
the singularity and the form factor come from pQCD), it can
manifest itself only in $\gamma^*\gamma^*$, but we have seen that there
is no definite need for such a singularity in present data.

\section*{Appendix 1: general proof of $t$-channel unitarity relations}

Assuming that $ m_{b} $ is the lowest hadronic mass, we know that
$aa\rightarrow aa$, $ab\rightarrow ab$ and $bb\rightarrow bb$
 have thresholds for $ t>4m_{a}^{2}>4m_{b}^{2} $.
In general, if $ t $ is large enough, there are many possible
intermediate states for each process under consideration,
which we must take
into account to write the unitarity relations. These states can be grouped
into subsets which have the same quantum numbers, and for which one can
derive factorisation.

Starting with the unitarity of the $ S $ matrix:
\begin{equation}
S^{\dagger }S=SS^{\dagger }=\1
\end{equation}
and setting $ S=\1+iS_{c} $, we obtain
\begin{equation}
\label{unita}
S_{c}-S_{c}^{\dagger }=iS_{c}^{\dagger }S_{c}=iS_{c}S_{c}^{\dagger }.
\end{equation}
One can define the invariant amplitude $ T_{if} $ by the matrix
elements\begin{equation}
<f|S_{c}|i>=(2\pi )^{4}\delta ^{4}(p_{f}-p_{i})T_{if}.\end{equation}
Eq. (\ref{unita}) then becomes the following at the amplitude level:
\begin{equation}
T_{if}-T_{if}^{\dagger }=C_{s}(T,T^{\dagger
}).\label{unitarityy0}\end{equation}

We define the $ C_{s} $ operator as the following convolution:
\begin{equation}
\label{unitarity0}
C_{s}(T^{\dagger },T)=C_{s}(T,T^{\dagger })=2i\sum _{k}\int dPS\:
T_{ik}T^{\dagger }_{kf}
\end{equation}
where $ k $ labels the possible intermediate on-shell $n$-particle
states in the $ t $ channel, which can differ by the number and
nature of produced particles, and $ dPS $ represents the differential
$n$-particle Lorentz-invariant phase space associated with these states.

If the particles are massive, we can enumerate these open channels
and assume that $ k $ runs from 1 to $ N+2 $, the number of open channel
depending on the value of $t$, {\it i.e.} of the energy in that channel.
In particular, we
shall find in this set of states the $ a\overline{a} $ and $ b\overline{b} $
intermediate states to which we respectively assign the labels $ k=1,\: 2 $.
Note that in general the label $ k $ does not refer to the number
of particles in the intermediate state, and that $ k $ can stand
for particles different from $ a $ and $ b $. So in general
the amplitude $ T_{km} $ represents the following process:
\begin{center}
\epsfig{file=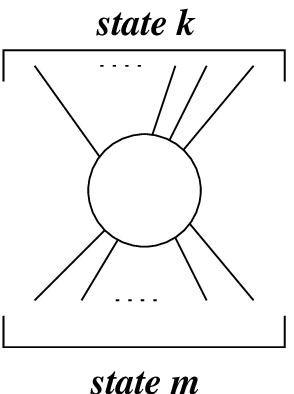}
\end{center}
Eq. (\ref{unitarityy0}) can then be represented by:
\begin{center}
\epsfig{file=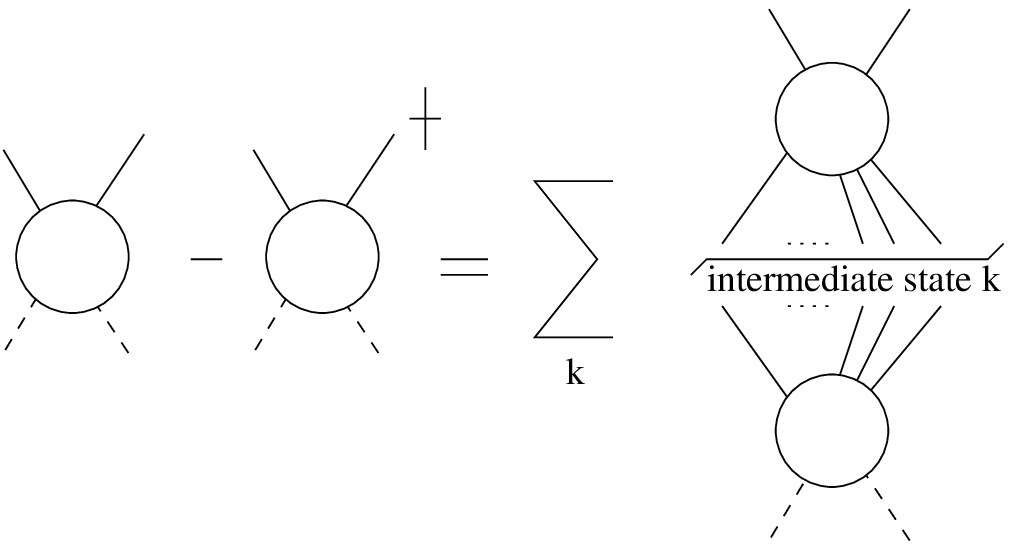,width=9cm}
\end{center}

We can now imagine that we split the amplitude into  charge-parity $+1 $
and  charge-parity $ -1 $ parts, and perform a Watson-Sommerfeld transform
\begin{equation}
\label{Regge}
T_{ab}^{\pm }(\nu ,t)=16\pi i\int dl\; P_{l}(-\cos (\vartheta _{t}))\;
\frac{2l+1}{2\sin (\pi l)}T_{ab}^{\pm }(l,t)\; \left( 1\pm e^{-i\pi l}\right) .
\end{equation}
with $\nu=p.q$.
(In the following, we shall only consider the charge-parity $+1 $ part of
the amplitudes without carrying the superscript $ + $.) After continuing
this relation to complex $ l\equiv j $, we deform the contour of integration
so that only the singularities of $ T(j,\, t) $ will contribute.
All amplitudes become functions of $ j $, and the operator
$ C_{s} $ changes to $ C $, which has the following properties:
\begin{itemize}
\item It is associative and distributive\begin{equation}
C(\alpha A_{1}+\beta A_{2},B)=\alpha C(A_{1},B)+\beta C(A_{2},B).\end{equation}
\item In the case of 2-particle intermediate states $ k $, the form of
$ C $ is particularly simple:\begin{equation}
C_{2}(T^{\dagger },T)=\rho _{k}T_{ik}T^{\dagger }_{kf}=(TRT^{\dagger
})_{if}\end{equation}
with $ \rho _{k}=2i\sqrt{\frac{t-4m_{k}^{2}}{t}} $, and $ R_{km}=\rho
_{k}\delta _{km} $.
\end{itemize}
To proceed further, we shall represent the $ T $ matrix in the
following form, for $ k\leq N+2 $:
\begin{eqnarray}
T=\left( \begin{array}{c|c}
T_{0}(2\times 2) &  T_{u}(2\times N)\\
\hline
T_{l}(N\times 2) &  T_{r}(N\times N)
\end{array}\right)
,\end{eqnarray}
where we have indicated the dimensions of the sub-matrices in parenthesis.
$ T_{0} $ contains the elastic amplitudes ($A^{(1)}_{if}$, $ i,\: f $=1, 2),
the upper matrix $ T_{u} $ contains the inelastic amplitudes $ i=1,2\rightarrow
k>2 $,
and the lower matrix $ T_{l} $ the inelastic amplitudes $ k>2\rightarrow i=1,\:
2 $.
$ T_{r} $ stands for the rest of the amplitudes $ k\rightarrow m $,
with $ k $ and $ m $ $ >2 $.

The system (\ref{unitarityy0}) can then be written:
\begin{eqnarray}
\label{first}
T_{0}-T_{0}^{\dagger }&=&T_{0}RT_{0}^{\dagger }+C(T_{u},T_{u}^{\dagger }),\\
\label{two}
T_{u}-T_{l}^{\dagger }&=&T_{0}RT_{l}^{\dagger }+C(T_{u},T_{r}^{\dagger }),\\
\label{three}
T_{l}-T_{u}^{\dagger }&=&T_{l}RT_{0}^{\dagger }+C(T_{r},T_{u}^{\dagger }),\\
\label{four}
T_{r}-T_{r}^{\dagger }&=&C(T_{l},T_{l}^{\dagger })+C(T_{r},T_{r}^{\dagger }).
\end{eqnarray}

To derive factorisation, it is enough to consider the first two relations
(\ref{first}, \ref{two}). We assume that the second equation can
be solved by a series expansion, yielding
\begin{equation}
T_{u}=M+T_{0}RM
\end{equation}
with $ M $ the solution of $ M=T_{l}^{\dagger }+C(M,T_{r}^{\dagger }) $:
\begin{equation}
M=T_{l}^{\dagger }+C(T_{l}^{\dagger },T_{r}^{\dagger })+C(C(T_{l}^{\dagger
},T_{r}^{\dagger }),T_{r}^{\dagger })+...\end{equation}
We can put this form into Eq.~(\ref{first}), which then gives
\begin{equation}
T_{0}(\1-RD)=D
\end{equation}
with\begin{equation}
D=\left[ T_{0}^{\dagger }+C(M,T_{u}^{\dagger })\right] .\end{equation}
Then we can repeat the argument given in the main text after Eq.~(\ref{solut}),
leading to the factorisation relation (\ref{facto}) near each singularity.

\section*{Appendix 2: unitarity constraints for off-shell photons}
\label{offsp}
The virtual photons must not be included in the intermediate states
of \break Eq.~(\ref{unitarityy0}). For hadronic final states, and in the
one-photon approximation, the photons can be thought of as external
particles which get inserted into the hadronic cross section.

In this case, we want to indicate explicitly whether the external
legs of the $ 2\rightarrow 2 $, $ 2\rightarrow n $ and $ n\rightarrow 2 $
amplitudes are off-shell or not. We introduce the notations $
T_{0}(Q_{in},Q_{out}) $,
$ T_{u}(Q_{in}) $ and $ T_{l}(Q_{out}) $, where $ Q_{in} $
stands for the two virtualities $ (Q_1^{2},Q_2^{2}) $ of the
initial states in the $ t $ channel, and $ Q_{out} $ for the
two virtualities $ (Q_{3}^{2},Q_{4}^{2}) $ of the final states,
and we write $ Q_{in}=0 $ or $ Q_{out}=0 $ in the case of on-shell
states, and the relations (\ref{unitarityy0}) can be visualised as
follows:

\vspace{0.375cm}
{\centering \includegraphics{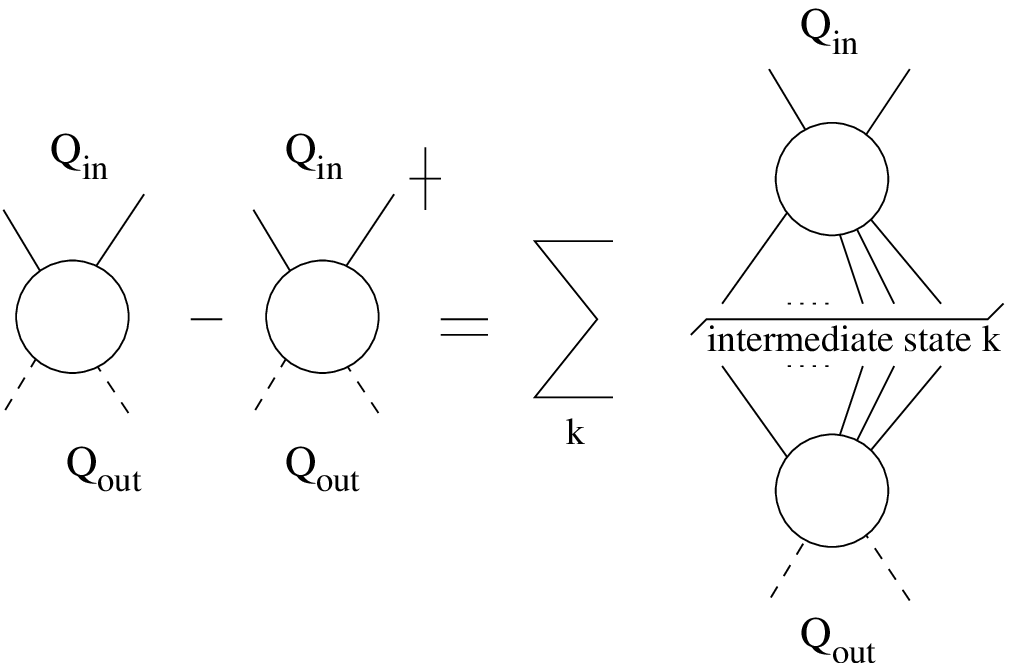} \par}
\vspace{0.375cm}

The system of equations (\ref{first}-\ref{four}) then becomes:

\begin{eqnarray}
T_{0}(Q_{in},Q_{out})-T_{0}^{\dagger
}(Q_{in},Q_{out})&=&T_{0}(Q_{in},0)RT_{0}^{\dagger }(0,Q_{out})\nonumber\\
&+&C(T_{u}(Q_{in}),T_{u}^{\dagger }(Q_{out})),
\label{First}\\
\label{Two}
T_{u}(Q_{in})-T_{l}^{\dagger }(Q_{in})&=&T_{0}(Q_{in},0)RT_{l}^{\dagger
}(0)+C(T_{u}(Q_{in}),T_{r}^{\dagger }),\\
\label{Three}
T_{l}(Q_{out})-T_{u}^{\dagger }(Q_{out})&=&T_{l}(Q_{out})RT_{0}^{\dagger
}(0,0)+C(T_{r},T_{u}^{\dagger }),\\
\label{Four}
T_{r}-T_{r}^{\dagger }&=&C(T_{l}(0),T_{l}^{\dagger }(0))+C(T_{r},T_{r}^{\dagger
}).
\end{eqnarray}

The resolution of the system proceeds as in Appendix 1 with the elimination
of\break $T_{u}(Q_{in}) $:
\begin{equation}
T_{u}(Q_{in})=M(Q_{in})+T_{0}(Q_{in},0)RM(0)\end{equation}
with $ M $ the solution of $ M(Q)=T_{l}^{\dagger }(Q)+C(M(Q),T_{r}^{\dagger })
$:\begin{equation}
M(Q_{in})=T_{l}^{\dagger }(Q_{in})+C(T_{l}^{\dagger }(Q_{in}),T_{r}^{\dagger
})+C(T_{l}^{\dagger }(Q_{in}),T_{r}^{\dagger }),T_{r}^{\dagger
})+...\end{equation}

The first equation however now gives
\begin{equation}
\label{Master}
T_{0}(Q_{in},Q_{out})=D(Q_{in},Q_{out})+T_{0}(Q_{in},0)RD(0,Q_{out})
\end{equation}
with \begin{equation}
D(Q_{in},Q_{out})=T_{0}^{\dagger }(Q_{in},Q_{out})+C(M(Q_{in}),T_{u}^{\dagger
}(Q_{out})).\end{equation}

For DIS, we consider $ Q_{out}=0 $ and $ Q_{1}^{2}=Q_{2}^{2}=Q^{2}\equiv -q^2
$.
(Note that the same kind of relations and conclusions would hold for
off-forward parton distribution functions). This gives us
\begin{equation}
T_{0}(Q_{in},0)(\1-RD(0,0))=D(Q_{in},0).\end{equation}

Hence we see that all the on-shell singularities must be present in
the off-shell case, due to the factor $(\1-RD(0,0))$,
 but we can have new ones coming from the singularities
of $ D(Q_{in},0) $. These singularities can be of perturbative
origin (\emph{e.g.} the singularities generated by the DGLAP evolution)
but their coupling will depend on the threshold matrix $ R $, and
hence they must know about hadronic masses, or in other words they
are not directly accessible by perturbation theory.

In the case of $ \gamma ^{*}\gamma ^{*} $ scattering, we take $Q_{in}=Q^2$
and $Q_{out}=P^2=-p^2$, and Eq.~(\ref{Master})
gives
\begin{equation}
\label{Master2b}
T_{0}(Q_{in},Q_{out})=D(Q_{in},Q_{out})
+\frac{D(Q_{in},0)RD(0,Q_{out})}{\1-RD(0,0)}.
\end{equation}
This shows that the DIS singularities will again be present, either through
$ \Delta  $, or through extra singularities present in DIS (in
which case their order will be different in $ \gamma \gamma  $
scattering, at least for $Q_{in}=Q_{out}$).

It is also possible to have extra singularities purely from $ D(Q_{in},Q_{out})
$.
\emph{A priori} these could be independent from the threshold matrix
$R$,
and hence be of purely perturbative origin
(\emph{e.g.} $ \gamma ^{*}\gamma ^{*}\rightarrow \bar{q}q $
or the BFKL pomeron coupled to photons through a perturbative impact factor).

\section*{Appendix 3: the box diagram}

We have re-calculated the contribution of the box diagram of
Fig.~(\ref{boxdi}),
and confirm the results of \cite{Budnev}. Our results can be recast in the
following form, which may be more transparent
in the present context:

We use $x_1=P^2/(2\nu)$ and $x_2=Q^2/(2\nu)$, with $\nu=p.q$, which give
\begin{eqnarray}
P^2&=&{x_1 w^2\over 1-x_1-x_2},\\
Q^2&=&{x_2 w^2\over 1-x_1-x_2}\end{eqnarray}
with $w^2=s$.
We set
\begin{eqnarray} \mu&=&{m^2\over\nu}={2 m^2(1-x_1-x_2)\over w^2},\\
{\tau}&=&{1-4 x_1 x_2},\\
\delta&=&{-x_1-x_2+1},\\
\delta_\mu&=&\delta-2\mu.\end{eqnarray}
The cross sections then take the form
\begin{equation}\sigma_i={12\alpha^2 \pi\delta\over
w^2}\left[{\sqrt{\delta_\mu}\over \sqrt{\delta \tau} (2\delta x_1
x_2+\tau \mu) \tau^2 }{\Sigma}_i+{\Lambda_i\over\tau^3}
 \log(\rho)\right]\end{equation}
which gives
\begin{equation}\rho={\sqrt{\delta\delta_\mu\tau}-\delta_\mu \tau
\over \sqrt{\delta\delta_\mu\tau}+\delta_\mu \tau
}.\end{equation}

The cross sections then are built from:
\begin{eqnarray}{\Sigma}_{TT}&=&
4 \delta x_1 x_2 [2 x_1 x_2 (x_1^2+x_2^2-1+2 x_1+2x_2)
\nonumber\\
&-&12 x_1^2 x_2^2
-2 (x_1^2+x_2^2)+2 (x_2+x_1)
-1] \nonumber\\
&-&\tau\mu (2 x_1-1)^2 (2 x_2-1)^2  -2 \delta\mu^2 \tau^2, \nonumber\\
\Lambda_{TT}&=&2 \delta \mu \tau-2 \mu^2 \tau^2\nonumber\\
&+&[8 x_1^2x_2^2(x_1^2+x_2^2)+16 x_1^3 x_2^3
-16 x_1^2 x_2^2(x_1+x_2)\nonumber\\
&-&4 x_1x_2(x1^2+x_2^2)
+16 x_1^2 x_2^2+
-2 (x_1+x_2)+2 (x_1^2+x_2^2)+1],\nonumber\\
{\Sigma}_{TL}&=&\mu \tau \delta x_2 [(6 x_1^2+1+2 x_2x_1-6 x_1) \nonumber\\
&+&2 \delta x_1 ((2 x_1^2+1) x_2+(2 x_2^2
+1) x_1-6x_1x_2) ], \nonumber\\
\Lambda_{TL}&=&- x_2[2\delta x_1 (2 x_2-1-2 x_1^2 x_2-2x_1x_2^2-2x_1x_2+2x_1)
\nonumber\\
&+&\mu \tau(2 x_1^2+1-2 x_2x_1+x_1)],\nonumber\\
{\Sigma}_{LT}&=&{\Sigma}_{TL}(x_1\leftrightarrow x_2),\nonumber\\
\Lambda_{LT}&=&\Lambda_{TL}(x_1\leftrightarrow x_2),\nonumber\\
{\Sigma}_{LL}&=&-2 \delta^2 x_1 x_2(3 \delta x_1 x_2+\mu \tau), \nonumber\\
\Lambda_{LL}&=&- \delta^2 x_1 x_2 (2 x_1 x_2+1).\nonumber
\end{eqnarray}

\section[*]{Acknowledgements} J.R.C. acknowledges the contribution
of P.V. Landshoff
who initiated this research by suggesting the possibility of extra
singularities from $t$-channel unitarity, G.S. is supported as Aspirant du
Fonds National pour la Recherche Scientifique, Belgium,
which also supports E.M. as a Visiting Fellow.

\end{document}